\documentclass[preprint]{aastex}

\newcommand{\yr}{{\rm\,yr}}

\newcommand{\km}{{\rm\,km}}

\begin{document}

\title{On the Orbits and Masses of the Satellites of the Pluto-Charon System}
\author{Man Hoi Lee and S. J. Peale}
\affil{Department of Physics, University of California,
       Santa Barbara, CA 93106}

\begin{abstract}
Two small satellites of Pluto, S/2005 P1 (hereafter P1) and S/2005 P2
(hereafter P2), have recently been discovered outside the orbit of
Charon, and their orbits are nearly circular and nearly coplanar with
that of Charon.
Because the mass ratio of Charon-Pluto is $\sim 0.1$, the orbits of P2
and P1 are significantly non-Keplerian even if P2 and P1 have
negligible masses.
We present an analytic theory, with P2 and P1 treated as test
particles, which shows that the motion can be represented by the
superposition of the circular motion of a guiding center, the forced
oscillations due to the non-axisymmetric components of the potential
rotating at the mean motion of Pluto-Charon, the epicyclic motion, and
the vertical motion.
The analytic theory shows that the azimuthal periods of P2 and P1
are shorter than the Keplerian orbital periods, and this deviation
from Kepler's third law is already detected in the unperturbed
Keplerian fit of Buie and coworkers.
In this analytic theory, the periapse and ascending node of each of
the small satellites precess at nearly equal rates in opposite
directions.

From direct numerical orbit integrations, we show the increasing
influence of the proximity of P2 and P1 to the 3:2 mean-motion
commensurability on their orbital motion as their masses increase
within the ranges allowed by the albedo uncertainties.
If the geometric albedos of P2 and P1 are high and of order of that of
Charon, the masses of P2 and P1 are sufficiently low that their orbits
are well described by the analytic theory.
The variation in the orbital radius of P2 due to the forced
oscillations is comparable in magnitude to that due to the best-fit
Keplerian eccentricity, and there is at present no evidence that P2
has any significant epicyclic eccentricity.
However, the orbit of P1 has a significant epicyclic eccentricity, and
the prograde precession of its longitude of periapse with a period of
$5300\,$days should be easily detectable.
If the albedos of P2 and P1 are as low as that of comets, the large
inferred masses induce significant short-term variations in the
epicyclic eccentricities and/or periapse longitudes on the
400--500-day timescales due to the proximity to the 3:2
commensurability.
In fact, for the maximum inferred masses, P2 and P1 may be in the 3:2
mean-motion resonance, with the resonance variable involving the
periapse longitude of P1 librating.
Observations that sample the orbits of P2 and P1 well on the
400--500-day timescales should provide strong constraints on the
masses of P2 and P1 in the near future.
\end{abstract}

\section{INTRODUCTION}

\citet{wea06} have recently discovered two new satellites of Pluto,
S/2005 P1 (hereafter P1) and S/2005 P2 (hereafter P2), from images
taken with the {\it Hubble Space Telescope} (HST).
These are the first new satellites of Pluto since the discovery of
Charon in 1978 \citep{chr78}.
The new satellites are much fainter and hence much smaller than
Charon (whose diameter $\sim 1200\km$; e.g., \citealt{sic06}), with
the diameter of P1 $\sim 60$--$170\,$km depending on the geometric
albedo, and P2 about $20\%$ smaller than P1.
Since the discovery data consist of only two epochs separated by three
days in May 2005, \citet{wea06} were unable to determine the orbits of
P2 and P1, but the data are consistent with the orbits of P2 and P1
being nearly circular and nearly coplanar with that of Pluto-Charon,
with orbital periods of $\sim 25\,$days for P2 and $\sim 38\,$days for
P1.
\citet{ste06} have used the same HST observations to place constraints
on the existence of any additional satellites of Pluto.

Prior to the discovery of P2 and P1, \citet[][hereafter BGYYS]{bui06}
have obtained HST images of the Pluto system in a series of 12 visits
from June 2002 to June 2003 for the purpose of producing an albedo map
of Pluto.
BGYYS were able to detect Charon in individual frames and P2 and P1 by
stacking the images taken during each visit.
BGYYS fit this data, together with the data for Charon from
\citet{tho97} and the data for P2 and P1 from \citet{wea06}, by
assuming that all three satellites are on unperturbed Keplerian
orbits.
The best-fit Keplerian orbital parameters found by BGYYS and their
1-$\sigma$ errors are reproduced in Table \ref{table1}.
The best-fit orbit of Charon relative to Pluto is consistent with zero
eccentricity, and BGYYS pointed out that the nonzero eccentricity
reported by \citet{tho97} is probably due to the use of an imprecise
center of body of Pluto in the earlier paper.
The parameters in Table \ref{table1} show that the orbits of P2 and P1
are indeed nearly circular and nearly coplanar with that of
Pluto-Charon.
Because P2 and P1 are much smaller than Pluto and Charon, they orbit
about a point that is very close to the center of mass of Pluto and
Charon.
Thus one of the fitting parameters is the mass ratio of Charon-Pluto,
and BGYYS found $m_c/m_p = 0.1165 \pm 0.0055$.

Two results from the orbit-fitting suggest that unperturbed Keplerian
orbits are not good assumptions for the orbits of P2 and P1.
The orbital period $P$ and semimajor axis $a$ of each orbit are
independent parameters in the fits by BGYYS.
If we assume that Kepler's third law is valid for all of the orbits,
each orbit yields an independent measurement of the total mass of
Pluto-Charon.
BGYYS found $m_p + m_c = 1.4570 \pm 0.0009 \times 10^{22}\,$kg from
Charon's orbit, $1.480 \pm 0.011 \times 10^{22}\,$kg from P2's orbit,
and $1.4765 \pm 0.006 \times 10^{22}\,$kg from P1's orbit.\footnote{
We note that the masses reported by BGYYS are consistent with using
$G = 6.67 \times 10^{-11}\,{\rm m}^3\,{\rm kg}^{-1}\,{\rm s}^{-2}$
instead of
$G = 6.672 \times 10^{-11}\,{\rm m}^3\,{\rm kg}^{-1}\,{\rm s}^{-2}$ in
converting from $G (m_p + m_c) = (2\pi/P)^2 a^3$ to $m_p + m_c$.
}
The values of $m_p + m_c$ from P2 and P1 disagree with that from
Charon at the 2.1-$\sigma$ and 3.2-$\sigma$ levels, respectively.
As we shall see, the discrepancies are in fact due to the deviation
from Kepler's third law for P2 and P1.
Because of the rather large mass ratio of Charon-Pluto, the deviations
of the gravitational potential from that of a point mass and,
consequently, of the orbits of P2 and P1 from Keplerian orbits are
nontrivial, even if P2 and P1 can be treated as test particles.

The second result from the orbit-fitting that suggests non-Keplerian
orbits for P2 and P1 is the confirmation of the result of
\citet{wea06} that the orbital periods of Charon, P2, and P1 are
nearly in the ratio 1:4:6.
This means that the orbits of P2 and P1 could be strongly affected by
resonant or near-resonant interactions.
As we shall see, the strongest effects come from the proximity of P2
and P1 to the 3:2 mean-motion commensurability, which is the lowest
order commensurability among the satellites, even though P2 and P1 are
much smaller than Charon.

In Section 2 we present an analytic theory for the orbits of P2 and P1
that is valid in the limit that the satellites have negligible masses
and can be treated as test particles.
It shows that the motion can be represented by the superposition of
the circular motion of a guiding center, the forced oscillations due
to the non-axisymmetric components of the potential rotating at the
mean motion of Pluto-Charon, the epicyclic motion, and the vertical
motion.
It also gives analytic results for the deviation from Kepler's third
law and the periapse and nodal precession rates.
In Section 3 we present direct numerical orbit integrations with
different assumed masses for P2 and P1 within the ranges allowed by
the uncertainties in the albedos.
The numerical results are compared to the analytic theory in Section
2, and the increasing importance of the proximity to the 3:2
commensurability with increasing masses is examined.
In fact, for the maximum masses corresponding to the lowest expected
albedos, P2 and P1 may be in the 3:2 mean-motion resonance, with the
resonance variable involving the periapse longitude of P1 librating.
In Section 4 we summarize our results and discuss the prospects for
detecting non-Keplerian behaviors and putting constraints on the
masses of P2 and P1 with existing and future observations.

\section{ANALYTIC THEORY}

In this section we develop an analytic theory for the orbits of the
satellites P2 and P1 that is valid in the limit that the satellites
have negligible masses and can be treated as test particles.
The latest orbital fit show that the orbit of Charon relative to Pluto
is consistent with zero eccentricity (see Table \ref{table1}).
Thus we assume that the orbit of Charon relative to Pluto is Keplerian
and circular, with semimajor axis $a_{pc}$ and mean motion (or
circular frequency) $n_{pc} = [G (m_p + m_c)/a_{pc}^3]^{1/2}$, where
$m_p$ and $m_c$ are the masses of Pluto and Charon, respectively.
In a cylindrical coordinate system with the origin at the center of
mass of the Pluto-Charon system and the $z = 0$ plane being the
orbital plane of Pluto-Charon, the positions of Charon and Pluto are
$\mbox{\boldmath $r$}_c = (a_c, \phi_c, 0)$, and
$\mbox{\boldmath $r$}_p = (a_p, \phi_c+\pi, 0)$, respectively, where
$a_p = a_{pc} m_c/(m_p+m_c)$, $a_c = a_{pc} m_p/(m_p+m_c)$, $\phi_c(t)
= n_{pc} t + \varphi_{pc}$, and $\varphi_{pc}$ is a constant.

\subsection{Potential}

Since the orbital radii of the satellites are much smaller than the
Hill radius ($\approx 8.0 \times 10^6\km$) of the Pluto-Charon system,
the perturbation from the Sun can be ignored and the gravitational
potential at $\mbox{\boldmath $r$} = (R, \phi, z)$ is
\begin{equation}
\Phi(\mbox{\boldmath $r$}) =
- {G m_p \over \left|\mbox{\boldmath $r$} - \mbox{\boldmath $r$}_p\right|}
- {G m_c \over \left|\mbox{\boldmath $r$} - \mbox{\boldmath $r$}_c\right|} .
\end{equation}
The orbits of the satellites are nearly coplanar with that of
Pluto-Charon, and we expand
$1/|\mbox{\boldmath $r$} - \mbox{\boldmath $r$}_c|$
in powers of $z$:
\begin{equation}
{1 \over \left|\mbox{\boldmath $r$} - \mbox{\boldmath $r$}_c\right|}
= {1 \over \left(\rho^2 + z^2\right)^{1/2}}
= {1 \over \rho} - {1 \over 2} {z^2 \over \rho^3} + \cdots ,
\end{equation}
where
\begin{equation}
\rho = \left[R^2 + a_c^2 - 2 R a_c \cos\left(\phi-\phi_c\right)\right]^{1/2} .
\end{equation}
By expressing the inverse powers of $\rho$ as cosine series using the
Laplace coefficients (see Eq. [6.62] of \citealt{mur99}), we obtain
\begin{equation}
{1 \over \left|\mbox{\boldmath $r$} - \mbox{\boldmath $r$}_c\right|}
= {1 \over 2 R} \sum_{k=0}^{\infty} (2-\delta_{k0})
  \left[b_{1/2}^k(\alpha_c) - {1 \over 2} \left(z \over R\right)^2
        b_{3/2}^k(\alpha_c) + \cdots\right] \cos k(\phi-\phi_c) ,
\end{equation}
where $\delta_{k0}$ is the Kronecker delta, $\alpha_c = a_c/R$, and
$b_s^k(\alpha_c)$ are the Laplace coefficients.
With a similar expression for
$1/|\mbox{\boldmath $r$} - \mbox{\boldmath $r$}_p|$,
the potential can be written as
\begin{equation}
\Phi(\mbox{\boldmath $r$}) =
  \sum_{k=0}^{\infty} \left[
  \Phi_{0k} (R) - {1 \over 2} \left(z \over R\right)^2 \Phi_{2k} (R) +
  \cdots \right] \cos k(\phi-\phi_c) ,
\label{Phi}
\end{equation}
where
\begin{equation}
\Phi_{jk} (R) = - {2-\delta_{k0} \over 2} \left[
                {m_c \over (m_p+m_c)} b_{(j+1)/2}^k(\alpha_c) + (-1)^k
                {m_p \over (m_p+m_c)} b_{(j+1)/2}^k(\alpha_p)
                \right] {G (m_p+m_c) \over R} .
\end{equation}
The axisymmetric $k=0$ components of the potential are identical to
those due to two rings --- one of mass $m_p$ and radius $a_p$ and
another of mass $m_c$ and radius $a_c$ --- at the $z=0$ plane.

Since $a_{pc}/R \approx 0.40$ and $0.30$ for P2 and P1, respectively,
it is instructive to examine the expansion of $\Phi_{0k}$ in powers of
$a_{pc}/R$ for the lowest values of $k$:
\begin{eqnarray}
\Phi_{00}
&=& - \Bigg[1 +
    {1 \over 4 (1+m_c/m_p)^2}
      \left(m_c \over m_p\right) \left(a_{pc} \over R\right)^2
\label{Phi00} \\
& & \qquad +
    {9 (1 - m_c/m_p + m_c^2/m_p^2) \over 64 (1+m_c/m_p)^4}
      \left(m_c \over m_p\right) \left(a_{pc} \over R\right)^4 + \cdots
    \Bigg] {G (m_p+m_c) \over R} ,
\nonumber \\
\Phi_{01}
&=& - \Bigg[
    {3 (1-m_c/m_p) \over 8 (1+m_c/m_p)^3}
      \left(m_c \over m_p\right) \left(a_{pc} \over R\right)^3 + \cdots
    \Bigg] {G (m_p+m_c) \over R} ,
\label{Phi01} \\
\Phi_{02}
&=& - \Bigg[
    {3 \over 4 (1+m_c/m_p)^2}
      \left(m_c \over m_p\right) \left(a_{pc} \over R\right)^2
\label{Phi02} \\
& & \qquad +
    {5 (1 - m_c/m_p + m_c^2/m_p^2) \over 16 (1+m_c/m_p)^4}
      \left(m_c \over m_p\right) \left(a_{pc} \over R\right)^4 + \cdots
    \Bigg] {G (m_p+m_c) \over R} ,
\nonumber \\
\Phi_{03}
&=& - \Bigg[
    {5 (1-m_c/m_p) \over 8 (1+m_c/m_p)^3}
      \left(m_c \over m_p\right) \left(a_{pc} \over R\right)^3 + \cdots
    \Bigg] {G (m_p+m_c) \over R} ,
\label{Phi03}
\end{eqnarray}
and also the expansion of $\Phi_{20}$:
\begin{eqnarray}
\Phi_{20}
&=& - \Bigg[1 +
    {9 \over 4 (1+m_c/m_p)^2}
      \left(m_c \over m_p\right) \left(a_{pc} \over R\right)^2
\label{Phi20} \\
& & \qquad +
    {225 (1 - m_c/m_p + m_c^2/m_p^2) \over 64 (1+m_c/m_p)^4}
      \left(m_c \over m_p\right) \left(a_{pc} \over R\right)^4 + \cdots
    \Bigg] {G (m_p+m_c) \over R} .
\nonumber
\end{eqnarray}
From the series definition of the Laplace coefficients (Eq. [6.68] of
\citealt{mur99}), one expects the lowest order terms of $\Phi_{jk}$ to
be of the order of $(a_{pc}/R)^k$.
However, the terms of the order of $a_{pc}/R$ in $\Phi_{01}$ cancel,
and the lowest order nonzero terms in $\Phi_{01}$ is of the order of
$(a_{pc}/R)^3$.
Eqs. (\ref{Phi00}) and (\ref{Phi20}) show that the axisymmetric
components of the potential deviate from that of a point mass of mass
$m_p + m_c$ at the origin by terms of the order of $(a_{pc}/R)^2$ and
higher, while Eqs. (\ref{Phi01})--(\ref{Phi03}) show that the
non-axisymmetric components of the potential are weak and of the order
of $(a_{pc}/R)^2$ and higher.
It should be noted that the deviations from a point mass potential are
multiplied by an additional small quantity $m_c/m_p$.

\subsection{Equations of Motion and Solution}

The equations of motion in cylindrical coordinates are
\begin{eqnarray}
\ddot{R} - R {\dot{\phi}}^2 &=&
     - {\partial \Phi \over \partial R} , \nonumber \\
R \ddot{\phi} + 2 \dot{R} \dot{\phi} &=&
     - {1 \over R}{\partial \Phi \over \partial \phi} , \label{EOM} \\
\ddot{z} &=&
     - {\partial \Phi \over \partial z} . \nonumber
\end{eqnarray}
The gravitational potential $\Phi$ (Eq. [\ref{Phi}]) is weakly
non-axisymmetric and rotates at a constant pattern speed $n_{pc}$,
and the satellites P2 and P1 are on nearly circular orbits that are
nearly coplanar with that of Pluto-Charon.
An approximate solution to the equations of motion that is valid for
such orbits can be obtained by representing the orbit as small
deviations from the circular motion of a guiding center in the $z=0$
plane:
\begin{eqnarray}
R &=& R_0 + R_1(t) , \nonumber \\
\phi &=& \phi_0(t) + \phi_1(t) , \label{Rphiz} \\
z &=& z_1(t) , \nonumber
\end{eqnarray}
where the constant $R_0$ is the radius of the guiding center,
$|R_1/R_0| \ll 1$, $|\phi_1| \ll 1$, and $|z_1/R_0| \ll 1$.
The approach is well known from the theory of orbits in weakly barred
galaxies (see, e.g., \citealt{bin87}) and planetary ring dynamics
(see, e.g., \citealt{gol82}).

Substituting Eqs. (\ref{Phi}) and (\ref{Rphiz}) into Eq.
(\ref{EOM}), the only nontrivial equation at the zeroth order is
\begin{equation}
R_0 {\dot{\phi_0}}^2 = \left[d\Phi_{00} \over dR\right]_{R_0} ,
\label{dotphi0}
\end{equation}
which describes the circular motion of the guiding center.
The solution of Eq. (\ref{dotphi0}) is
\begin{equation}
\phi_0(t) = n_0 t + \varphi_0 ,
\end{equation}
where $\varphi_0$ is a constant and the mean motion $n_0$ is given by
\begin{eqnarray}
n_0^2 
&=& \left[{1 \over R}{d\Phi_{00} \over dR}\right]_{R_0} \\
&=& {1 \over 2} \Bigg\{
    {m_c \over (m_p+m_c)} b_{1/2}^0(\alpha_c) +
    {m_p \over (m_p+m_c)} b_{1/2}^0(\alpha_p) \label{n0} \\
& & \qquad + {m_p m_c \over (m_p + m_c)^2} \left(a_{pc} \over R_0\right)
             \left[Db_{1/2}^0(\alpha_c) + Db_{1/2}^0(\alpha_p)
             \right]
\Bigg\} n_K^2 . \nonumber
\end{eqnarray}
In the above equation $n_K = [G (m_p+m_c)/R_0^3]^{1/2}$ is the
Keplerian mean motion at $R_0$, $D = d/d\alpha$, and $\alpha_c$ and
$\alpha_p$ are evaluated at $R = R_0$.

At the first order, the equation for $\phi_1$ is
\begin{equation}
\ddot{\phi}_1 + {2 n_0 \over R_0} \dot{R}_1
= \sum_{k=1}^{\infty} {k \Phi_{0k}(R_0) \over R_0^2} \sin(k\Delta\phi) ,
\label{ddotphi1}
\end{equation}
where
\begin{equation}
\Delta\phi
= \phi_0 - \phi_c
= (n_0 - n_{pc}) t + \varphi_0 - \varphi_{pc} .
\end{equation}
An integration of Eq. (\ref{ddotphi1}) yields an expression for
$\dot{\phi}_1$, which can then be substituted into the first order
equation for $R_1$ to yield
\begin{equation}
\ddot{R}_1 + \kappa_0^2 R_1 = - \sum_{k=1}^{\infty}
     \left[{d\Phi_{0k} \over dR} +
           {2 n \Phi_{0k} \over R (n - n_{pc})}
     \right]_{R_0} \cos(k\Delta\phi) ,
\label{ddotR1}
\end{equation}
where the epicyclic frequency $\kappa_0$ is given by
\begin{eqnarray}
\kappa_0^2
&=& \left[R{dn^2 \over dR} + 4n^2\right]_{R_0}
\label{kappa0def} \\
& & \nonumber \\
& & \nonumber \\
&=& {1 \over 2} \Bigg\{
    {m_c \over (m_p+m_c)} b_{1/2}^0(\alpha_c) +
    {m_p \over (m_p+m_c)} b_{1/2}^0(\alpha_p)
\nonumber \\
& & \qquad - {m_p m_c \over (m_p + m_c)^2} \left(a_{pc} \over R_0\right)
             \left[Db_{1/2}^0(\alpha_c) + Db_{1/2}^0(\alpha_p)
             \right]
\label{kappa0} \\
& & \qquad - {m_p m_c \over (m_p + m_c)^2} \left(a_{pc} \over R_0\right)^2
             \left[{m_p \over (m_p+m_c)} D^2b_{1/2}^0(\alpha_c) + {m_c
		 \over (m_p+m_c)} D^2b_{1/2}^0(\alpha_p)
             \right]
\Bigg\} n_K^2 .
\nonumber
\end{eqnarray}
In Eqs. (\ref{ddotR1}) and (\ref{kappa0def}),
$n = (R^{-1} d\Phi_{00}/dR)^{1/2}$ is the mean motion at $R$, and the
quantities in the square brackets are evaluated at $R = R_0$.
Eq. (\ref{ddotR1}) is the equation of motion of a simple harmonic
oscillator of natural frequency $\kappa_0$ that is driven at
frequencies $k|n_0-n_{pc}|$.
It is straightforward to solve Eqs. (\ref{ddotphi1}) and
(\ref{ddotR1}) to obtain
\begin{eqnarray}
R &=&
R_0 \left[1 - e \cos(\kappa_0 t + \psi) +
          \sum_{k=1}^{\infty} C_k \cos(k\Delta\phi)
    \right] , \label{Rt} \\
\phi &=&
n_0 t + \varphi_0 +
     {2 n_0 \over \kappa_0} e \sin(\kappa_0 t + \psi) -
     {n_0 \over (n_0 - n_{pc})} \sum_{k=1}^{\infty}
          {D_k \over k} \sin(k\Delta\phi) , \\
\dot{R} &=&
R_0 \left[e \kappa_0 \sin(\kappa_0 t + \psi) -
           (n_0 - n_{pc}) \sum_{k=1}^{\infty} k C_k \sin(k\Delta\phi)
    \right] , \label{dotRt} \\
\dot{\phi} &=&
n_0 \left[1 + 2 e \cos(\kappa_0 t + \psi) -
               \sum_{k=1}^{\infty} D_k \cos(k\Delta\phi)
         \right] , \label{dotphit}
\end{eqnarray}
where $e$ and $\psi$ are constants,
\begin{equation}
C_k = -\left[{1 \over R}{d\Phi_{0k} \over dR} +
             {2 n \Phi_{0k} \over R^2 (n - n_{pc})}
       \right]_{R_0} \Bigg/
       \left[\kappa_0^2 - k^2 (n_0 - n_{pc})^2
       \right] ,
\end{equation}
and
\begin{equation}
D_k = 2 C_k + {\Phi_{0k}(R_0) \over R_0^2 n_0 (n_0 - n_{pc})} .
\end{equation}
Finally, for the motion in $z$, the first order equation for $z_1$ is
\begin{equation}
\ddot{z}_1 + \nu_0^2 z_1 = 0 ,
\end{equation}
and its solution gives
\begin{equation}
z = z_1 = R_0 i \cos(\nu_0 t + \zeta)
\end{equation}
where $i$ and $\zeta$ are constants and the vertical frequency
$\nu_0$ is defined by
\begin{eqnarray}
\nu_0^2
&=& \left[-{\Phi_{20} \over R^2}\right]_{R_0}
\\
&=& {1 \over 2} \left[
    {m_c \over (m_p+m_c)} b_{3/2}^0(\alpha_c) +
    {m_p \over (m_p+m_c)} b_{3/2}^0(\alpha_p)
    \right] n_K^2 .
\label{nu0}
\end{eqnarray}

The motion in $R$ and $\phi$ is described by the superposition of
the circular motion of the guiding center at $R_0$ at frequency $n_0$,
the epicyclic motion represented by ``eccentricity'' $e$ at frequency
$\kappa_0$, and the forced oscillations of fractional radial
amplitudes $C_k$ at frequencies $k|n_0-n_{pc}|$.
The motion in $z$ decouples from that in $R$ and $\phi$ and has only
free oscillations at the vertical frequency $\nu_0$.
If there is no epicyclic or vertical motion ($e = i = 0$), the orbit
is a closed orbit in the frame rotating at the pattern speed $n_{pc}$,
with the variation of the radius with the azimuthal angle being the
sum of terms with fractional amplitude $C_k$ and period $2\pi/k$.
If $e \gg \sum_k C_k$, the orbit is approximately a precessing
Keplerian ellipse with eccentricity $e$, mean motion $n_0$, and
periapse precession rate
\begin{equation}
\dot{\varpi} = n_0 - \kappa_0 .
\end{equation}
If in addition $i \neq 0$, the orbit has an inclination $i$ with
respect to the Pluto-Charon orbital plane and a nodal precession rate
\begin{equation}
\dot{\Omega} = n_0 - \nu_0 .
\end{equation}

\subsection{Deviation from Kepler's Third Law}

Eq. (\ref{n0}) shows that the mean motion $n_0$ and hence the
azimuthal period $P_0 = 2\pi/n_0$ deviate from those of a Keplerian
orbit.
In Fig. \ref{fig:period} the solid line shows $P_K/P_0 - 1 =
n_0/n_K - 1$ from Eq. (\ref{n0}) as a function of the guiding
center radius $R_0$ for the best-fit mass ratio of Charon-Pluto,
$m_c/m_p = 0.1165$, while the dotted lines show the same quantity for
$m_c/m_p$ that are $1\,\sigma$ ($\pm 0.0055$) from the best-fit value.
(The uncertainty in $m_c/m_p$ dominates that in $a_{pc}$ in the
evaluation of Eq. [\ref{n0}].)
Fig. \ref{fig:period} shows that the azimuthal period is shorter
than the Keplerian orbital period.

The orbital parameters of P2 and P1 in Table \ref{table1} were
obtained by BGYYS from unperturbed Keplerian fits.
As we have shown in this section, the orbits of P2 and P1 are
non-Keplerian even if they can be treated as test particles.
Nevertheless, the orbital period $P$ and semimajor axis $a$ (which are
independent parameters in the fits by BGYYS) in Table \ref{table1}
should closely resemble the azimuthal period $P_0$ and guiding center
radius $R_0$.
With $P_K = 2\pi [a^3/G (m_p+m_c)]^{1/2} = P_{pc} (a/a_{pc})^{3/2}$,
where $P_{pc}$ and $a_{pc}$ are the orbital period and semimajor axis
of Pluto-Charon from Table \ref{table1}, we find $P_K/P - 1 = 0.0079
\pm 0.0038$ and $0.0067 \pm 0.0021$ for P2 and P1, respectively.
The values of $P_K/P - 1$ and $a$ for P2 and P1 are shown in Fig.
\ref{fig:period} with their $1\,\sigma$ error bars.
The orbital periods are clearly shorter than the Keplerian values (by
$2.1\,\sigma$ and $3.2\,\sigma$ for P2 and P1, respectively).
On the other hand, $P_K/P - 1$ for P2 is in excellent agreement
(within $0.4\,\sigma$) with the analytic result, and that for P1 is in
reasonable agreement (within $1.6\,\sigma$) with the analytic result.
The remaining discrepancy for P1 may be simply statistical, but it
could also be due to the assumption in the fitting that the orbit is
an unperturbed Keplerian orbit (see below for more details on the
expected non-Keplerian behaviors).

\subsection{Periapse and Nodal Precessions}

With $a$ from Table \ref{table1} as $R_0$ for P2 and P1 and $m_c/m_p =
0.1165$, we evaluate $P_K = P_{pc} (a/a_{pc})^{3/2}$, $n_0/n_K$
(Eq. [\ref{n0}]), $\kappa_0/n_K$ (Eq. [\ref{kappa0}]), and $\nu_0/n_K$
(Eq. [\ref{nu0}]) for the analytic theory, and they are listed in
Table \ref{table2}.
The precession of the periapse is prograde with period
$2\pi/|\dot{\varpi}| = 2\pi/|n_0 - \kappa_0| = 1740$ and $5280\,$days
for P2 and P1, respectively.
The nodal precession has a similar period ($2\pi/|\dot{\Omega}| =
2\pi/|n_0 - \nu_0| = 1770$ and $5330\,$days for P2 and P1,
respectively) but it is retrograde.

The periapse and nodal precessions at nearly equal rates in opposite
directions and the faster-than-Keplerian mean motion are similar to
the behaviors of orbits around an oblate planet (see, e.g., Section
6.11 of \citealt{mur99}).
This can be understood from the fact that the $(a_{pc}/R)^2$ terms in
the axisymmetric components of the potential, $\Phi_{00}$ and
$\Phi_{20}$ in Eqs. (\ref{Phi00}) and (\ref{Phi20}), are
identical to the $J_2$ terms of an oblate planet with $J_2 =
m_p m_c/[2 (m_p + m_c)^2]$.

\section{NUMERICAL ORBIT INTEGRATIONS}

\subsection{Initial Conditions and Numerical Methods}

For the remainder of this paper we use Jacobi coordinates where the
position of Charon is relative to Pluto, the position of the inner
satellite P2 is relative to the center of mass of Pluto-Charon, and
the position of the outer satellite P1 is relative to the center of
mass of Pluto-Charon-P2.
Jacobi coordinates are the natural generalization of the coordinates
used in Section 2 (where the position of P1 is relative to the center
of mass of Pluto-Charon) when P2 and P1 are not test particles, and
they reduce to the coordinates used in Section 2 in the test-particle
limit.

From $P_{pc}$ and $a_{pc}$ in Table \ref{table1}, we adopt
$G (m_p + m_c) = (2\pi/P_{pc})^2 a_{pc}^3 =
9.71791 \times 10^{11}\,{\rm m}^3\,{\rm s}^{-2}$
(or $m_p + m_c = 1.4565 \times 10^{22}\,$kg for
$G = 6.672 \times 10^{-11}\,{\rm m}^3\,{\rm kg}^{-1}\,{\rm s}^{-2}$).
For the mass ratio $m_c/m_p$, we use the best-fit value $0.1165$
from BGYYS.
We generate the initial position and velocity of Charon relative to
Pluto by using the orbital parameters in Table \ref{table1} at epoch
JD 2452600.5 as the osculating Keplerian orbital parameters.

The orbits of P2 and P1 are sufficiently non-Keplerian even in the
test-particle limit that, if we had generated their initial conditions
by assuming that the orbital parameters in Table \ref{table1} are the
osculating Keplerian parameters at epoch JD 2452600.5, the numerically
integrated orbits would have properties significantly different from
those of the best-fit Keplerian orbits in Table \ref{table1}, with the
mean orbital radii smaller than the semimajor axes in Table
\ref{table1} and the variations in the orbital radii larger than those
for the eccentricities in Table \ref{table1}.
Without refitting the data, we aim to adopt a set of initial
conditions so that the resulting orbits would closely resemble the
best-fit Keplerian orbits in Table \ref{table1}.
This is accomplished by using the mean longitudes $\lambda$ of P2 and
P1 in Table \ref{table1} as the initial values of $\phi$ and then
using Eqs. (\ref{Rt}), (\ref{dotRt}), and (\ref{dotphit}) to set
the initial values of $R$, $\dot{R}$, and $\dot{\phi}$.
As in Sections 2.3--2.4, we adopt the semimajor axes $a$ from the
Keplerian fit as the guiding center radii $R_0$.
We include the forced oscillation terms up to $k=4$, and the
coefficients $C_k$ and $D_k$ are listed in Table \ref{table2}, along
with $P_K$, $n_0/n_K$, $\kappa_0/n_K$, and $\nu_0/n_K$.
For the initial values of $\Delta\phi$ in the forced oscillation
terms, we can ignore the difference between $\phi$ and $\phi_0$ (with
the latter for the guiding center) and use $\phi - \phi_c$.

We can see from Table \ref{table2} that $\sum_k C_k \approx 0.0029$
for P2.
Thus the fractional orbital radius variation due to the forced
oscillation terms alone is comparable to that due to the best-fit
Keplerian eccentricity ($0.0023$), which is itself consistent with
zero ($\pm 0.0021$).
Therefore, we set the initial epicyclic eccentricity $e = 0$ for P2.
For P1, $\sum_k C_k \approx 0.0004$, which is much smaller than the
best-fit Keplerian eccentricity of $0.0052$($\pm 0.0011$), and we
adopt an initial epicyclic eccentricity $e = 0.0052$.
In the epicyclic approximation for a Keplerian orbit, the phase
$\kappa_0 t + \psi$ is the mean anomaly.
So it is reasonable to adopt for P1 $\psi = \lambda - \varpi$, where
$\lambda$ and $\varpi$ are the mean longitude and longitude of
periapse from Table \ref{table1}.

BGYYS have computed the 1-$\sigma$ contours of the orbit poles on the
J2000 sky plane for their best-fit Keplerian orbits for Charon, P2,
and P1 using the Monte Carlo technique, and the contours are shown in
their Fig. 2.
The 1-$\sigma$ contour for Charon is significantly smaller than those
for P2 and P1, with the latter two having mean radii of about
$0\fdg34$ and $0\fdg16$, respectively.
The orbit pole of Charon is $0\fdg10$ from that of P2 for the best-fit
orbits, which is well within the 1-$\sigma$ contour for P2, and it is
$0\fdg25$ from that of P1 for the best-fit orbits, which is about
$50\%$ further than the 1-$\sigma$ contour for P1 and only marginally
significant.
Since there is no significant detection of any mutual orbital
inclinations, we assume that the orbits of P2 and P1 are coplanar with
that of Pluto-Charon.
This means that the precession of nodes is not examined by our
numerical orbit integrations.

We perform 5 sets of integrations with different assumed masses for P2
and P1.
From the photometry of P2 and P1, \citet{wea06} have estimated that
the diameters of P2 and P1 are $46 \pm 4\km$ and $61 \pm 4\km$,
respectively, if the geometric albedos are Charon-like and $= 0.35$.
On the other hand, if the albedos are comet-like and $= 0.04$, the
diameters are $137 \pm 11\km$ for P2 and $167 \pm 10\km$ for P1.
If we assume that the mean density is $2{\rm\,g\,cm}^{-3}$ (i.e.,
similar to that of Pluto), $m_2 = 1.02 \times 10^{17}\,$kg and $m_1 =
2.38 \times 10^{17}\,$kg in the high albedo case, and $m_2 = 2.69
\times 10^{18}\,$kg and $m_1 = 4.88 \times 10^{18}\,$kg in the low
albedo case.
In addition to two integrations with the high- and low-albedo masses,
we perform an integration with masses $10^{-5}$ times those of the
high albedo case (so that P2 and P1 are test particles), an
integration with masses twice those of the high albedo case, and an
integration with masses half those of the low albedo case.

The direct numerical orbit integrations are performed using a
modified version of the \citet{wis91} symplectic integrator contained
in the SWIFT\footnote{
See \anchor{http://www.boulder.swri.edu/~hal/swift.html}
{http://www.boulder.swri.edu/$\sim$hal/swift.html}.
}
software package.
The Wisdom-Holman integrator is based on dividing the Hamiltonian of
the gravitational $N$-body problem into a part that describes the
Keplerian motions of the satellites around the planet (or, in the case
of a planetary system, the planets around the star) and a part that
describes the perturbations to the Keplerian motions.
The division used by \citet{wis91} assumes that all of the satellite
masses are much smaller than the planet mass, but in the case of
Pluto, $m_c/m_p = 0.1165$.
We have described in \citet{lee03} a modified Wisdom-Holman integrator
using a slightly different division of the Hamiltonian into the
Keplerian and perturbation parts.
The modified integrator is designed for hierarchical systems, where
the masses of the satellites can be comparable to that of the planet
but the orbit of each satellite is much larger than that of the
satellite just inside, and it was used by \citet{lee03} to study
hierarchical extrasolar planetary systems.
This modified integrator can also handle the Pluto system, where
$(m_c/m_p) (a_{pc}/R)^2 \la 0.019$, without an excessively small
timestep.
The integrations are performed with a timestep of $10^4\,$s (or about
55 steps per Charon's orbit).

\subsection{Results}

The numerical orbit integrations with test-particle masses,
high-albedo masses, and twice the high-albedo masses for P2 and P1
show that the orbits of P2 and P1 are well described by the analytic
theory in Section 2 if the masses of P2 and P1 are of the order of the
high albedo ones.
In Fig. \ref{fig:highR} we plot the variations in the orbital radii
$R_2$ and $R_1$ of P2 and P1, respectively, for $800\,$days in the
high albedo case.
The dashed lines indicate the semimajor axes, $a$, and the maximum and
minimum radii, $a (1 \pm e)$, for the best-fit Keplerian orbits in
Table \ref{table1}.
The orbital radius $R_2$ of P2 clearly shows the forced oscillations,
which are dominated by the $C_1$ and $C_2$ terms (see Table
\ref{table2}) with periods $2\pi/|n_0 - n_{pc}| \approx 8.6\,$days and
$\pi/|n_0 - n_{pc}| \approx 4.3\,$days, respectively.
As expected, the variation in $R_2$ due to the forced oscillations
is comparable in magnitude to that due to the best-fit Keplerian
eccentricity.
Although the initial epicyclic eccentricity of P2 is set to zero, the
fact that the forced oscillation terms with $k > 4$ are not included
in setting the initial conditions results in a small epicyclic motion
with a period $2\pi/\kappa_0 \approx 25.2\,$days, which is also
visible in the plot of $R_2$ in Fig. \ref{fig:highR}.
The variation in $R_1$ is dominated by the epicyclic motion with $e
\approx 0.0052$ and period $2\pi/\kappa_0 \approx 38.6\,$days.
For P1, the amplitudes of the forced oscillations are significantly
smaller than for P2, as we expect from the values of $C_k$ in Table
\ref{table2}, and the dominant forced oscillation periods are
$2\pi/|n_0 - n_{pc}| \approx 7.7\,$days and $\pi/|n_0 - n_{pc}|
\approx 3.8\,$days.
The $k > 4$ terms are sufficiently small for P1 that their neglect in
the initial conditions do not result in any significant additional
epicyclic motion.

In order to study in more detail the epicyclic component of the
motion, we need to eliminate the high frequency forced oscillations.
We find that the forced oscillations are sufficiently close to those
predicted by the analytic theory in all of our numerical integrations
that they can be effectively eliminated by defining a transformed
orbital radius:
\begin{equation}
R' = R - R_0 \sum_k C_k \cos[k(\phi - \phi_c)] ,
\label{Rprime}
\end{equation}
with $R_0$ and $C_k$ from Table \ref{table2} and $\phi$ and $\phi_c$
from the numerical integrations themselves (compare to
Eq. [\ref{Rt}]).
In Fig. \ref{fig:highRprime} we plot the variations in $R_2'$ and
$R_1'$ in the high albedo case, and it is clear from a comparison with
Fig. \ref{fig:highR} that most of the forced oscillations are
eliminated.
Fig. \ref{fig:highRprime} shows that there are small periodic
variations in the maximum and minimum values of $R_2'$ and $R_1'$, or
equivalently in the epicyclic eccentricities $e_2$ and $e_1$, in this
run with the high-albedo masses.
The amplitudes of the eccentricity variations are larger in the run
with twice the high-albedo masses, while the eccentricities are
nearly constant in the run with test-particle masses, which indicate
that the variations are due to interactions between P2 and P1 (see
below for more details).

The azimuthal period $P_0$ can be determined for the numerical
integrations from the cumulative increase in $\phi$.
We find $P_0 = 24.913$ and $38.335\,$days for P2 and P1, respectively,
for the run with high-albedo masses, and they are identical to $P_0$
found for the run with test-particle masses.\footnote{
It should be recalled that we adopt the best-fit semi-major axes $a$
from the Keplerian fit as the guiding center radii $R_0$ and that the
1-$\sigma$ error in $a$ is $\pm 121\km$ for P2 and $\pm 88\km$ for P1
(see Table \ref{table1}).
If we vary the guiding center radii by $\pm 1\,\sigma$ in the initial
conditions of our numerical integrations, we would find
$P_0 = 24.913 \pm 0.093$ and $38.335 \pm 0.078\,$days for P2 and P1,
respectively.
Thus, similar to the analytic results in Section 2.3, the azimuthal
periods from the numerical integrations agree with the best-fit
orbital periods in Table \ref{table1} from the Keplerian fit to within
$0.6\,\sigma$ for P2 and $1.6\,\sigma$ for P1.
}
If we use $P_K$ from Table \ref{table2}, $n_0/n_K = P_K/P_0 = 1.0056$
and $1.0033$ for P2 and P1, respectively.
The latter is in excellent agreement with the analytic value in Table
\ref{table2}, but the former is slightly smaller than the analytic
value.
The small discrepancy for P2 is due to the fact that the numerically
integrated orbit has a non-zero epicyclic eccentricity, with P2 near
the periapse at $t = 0$ (see Fig. \ref{fig:highRprime}), which means
that the guiding center radius of the numerically integrated orbit
($R_0 = 48698\km$ from the average of the maximum and minimum values
of $R_2'$) is slightly larger than the value we were aiming for ($R_0
= 48675\km$).
(For P1, the guiding center radius of the numerically integrated orbit
is identical to the value we were aiming for.)
If we use $P_K = 25.0695\,$days for the numerically determined $R_0$
instead of $P_K = 25.0518\,$days from Table \ref{table2} for P2, we
find $n_0/n_K = P_K/P_0 = 1.0063$, which is in excellent agreement
with the analytic value in Table \ref{table2} (note that the ratio
$n_0/n_K$ is much less sensitive to a small change in $R_0$ than $P_K$
and $P_0$ separately).

The easiest way to determine the longitude of periapse $\varpi$ of the
epicyclic motion as a function of time is to monitor the transformed
orbital radius $R'$ (Eq. [\ref{Rprime}]) during the numerical orbit
integration.
When $R'$ changes from decreasing in a previous timestep to increasing
in the current timestep, the satellite has passed the periapse, and we
can use the values of $R'$ and $\phi$ at the end of the current step
and two previous steps to find the time of periapse passage and
$\varpi$ ($=\phi$ at the time of periapse passage) by interpolation.
Similarly, the time at which $\varpi = \phi + 180^\circ$ (i.e.,
apoapse passage) can be determined.
Fig. \ref{fig:highvarpi} shows the evolution of $\varpi_2$ and
$\varpi_1$ of P2 and P1, respectively, for $10^4\,$days in the high
albedo case.
There are some spurious points in the plot of $\varpi_2$, because the
forced oscillations are not perfectly eliminated in the transformed
orbital radius and there are occasional false minima and maxima in
$R_2'$ when the epicyclic eccentricity $e_2$ is very small (see
Fig. \ref{fig:highRprime}).
In this high albedo case, the long-term evolution of $\varpi_2$ and
$\varpi_1$ are prograde precessions with periods of $2000$ and
$5300\,$days, respectively.
The latter is in excellent agreement with the analytic result in
Section 2.4, but the former is about $15\%$ longer.
The $15\%$ discrepancy for P2 is too large to be explained by the
slightly larger guiding center radius mentioned in the previous
paragraph, and it is not due to interactions between P2 and P1, as the
test-particle run shows the same precession period.
We suspect that the discrepancy is due to the neglect of second (and
higher) order corrections to the deviations from the guiding center
motion in the analytic theory in Section 2.

Superposed on the long-term periapse precessions are periodic
variations on the same timescales ($\approx 400$ and $450\,$days for
P2 and P1, respectively) as the epicyclic eccentricity variations seen
in Fig. \ref{fig:highRprime}.
For P2, the amplitude of the short-term variation is sufficiently
large that the precession is retrograde when the eccentricity is large
and prograde and faster than the long-term rate when the eccentricity
is small.
The short-term periodic variations in the epicyclic eccentricities and
periapse longitudes are due to interactions between P2 and P1, and we
can explain the periods by noting that their orbits are close to the
3:2 mean-motion commensurability (azimuthal period ratio $= 1.539$ for
the numerical integration, in agreement with the ratio 1.537 for the
best-fit orbital periods in Table \ref{table1}).
The disturbing potential for the interactions between P2 and P1, which
is not included in the analytic theory in Section 2, can be expanded
into a cosine series in the usual manner (see, e.g., \citealt{mur99}),
and because of the proximity to the 3:2 commensurability, we expect
the interactions to be dominated by the terms in the disturbing
potential associated with the cosine arguments (or resonance
variables) $\theta_2 = 2\phi_2 - 3\phi_1 + \varpi_2$ and  $\theta_1 =
2\phi_2 - 3\phi_1 + \varpi_1$.\footnote{
For convenience, we define the resonance variables $\theta_2$ and
$\theta_1$ using the true longitudes $\phi_2$ and $\phi_1$ instead of
mean longitudes.
While it is possible to define a mean longitude via a mean anomaly
that is proportional to $n_0 t$, it cannot be easily computed from the
position and velocity.
}
Fig. \ref{fig:highresvar} shows the evolution of $\theta_2$ and
$\theta_1$ for $10^4\,$days in the high albedo case.
It is clear from Fig. \ref{fig:highresvar} that neither $\theta_2$
nor $\theta_1$ is in resonance and that the circulation periods of
$\theta_2$ and $\theta_1$ are $\approx 400$ and $450\,$days,
respectively.
Thus the short-term periodic variations in $e_2$ and $\varpi_2$ are
associated with the circulation of $\theta_2$ and those in $e_1$ and
$\varpi_1$ are associated with the circulation of $\theta_1$.

The amplitudes of the short-term variations in $e$ and $\varpi$ are
larger for {\it both} P2 and P1 for larger masses of P2 and P1, as
long as the masses do not exceed about half the low-albedo ones.
In the run with half the low-albedo masses (Figs. \ref{fig:halflowR}
and \ref{fig:halflowvarpi}), $e_1$ and $\varpi_1$ show significant
variations, with the minimum $e_1$ much smaller than the initial $e_1
= 0.0052$.
Both $\varpi_2$ and $\varpi_1$ alternate between nearly linear
retrograde precession when the eccentricities are large and very fast
prograde precession when the eccentricities are small.
However, the long-term trends of both $\varpi_2$ and $\varpi_1$ remain
prograde, with periods of $1900$ and $5100\,$days, respectively, which
are slightly shorter than in the case with the high-albedo masses.

In the run with the low-albedo masses (Figs. \ref{fig:lowR} and
\ref{fig:lowvarpi}), while the properties of the orbit of P2 follow
the same trends with increasing masses as above (i.e., larger
amplitudes for the short-term variations of $e_2$ and $\varpi_2$ and a
shorter period of $1800\,$days for the long-term prograde precession
of $\varpi_2$), those of P1 do not follow the previous trends.
The amplitude of the variation in $e_1$ is now smaller than in the
case with half the low-albedo masses, and $\varpi_1$ shows a
completely retrograde precession with a period of $500\,$days.
The explanation can be found in the plot of the evolution of the
resonance variables $\theta_2$ and $\theta_1$ in Fig.
\ref{fig:lowresvar}.
While $\theta_2$ circulates as in the high albedo case shown in Fig.
\ref{fig:highresvar} (the nearly empty region about $180^\circ$ is due
to the fast prograde precession of $\varpi_2$ when $e_2$ is small),
$\theta_1$ librates about $180^\circ$.
Thus the case with the low-albedo masses differs from the cases with
lower masses in having $\theta_1$ in resonance.

\section{DISCUSSION AND CONCLUSIONS}

\begin{figure}
\epsscale{0.54}
\plotone{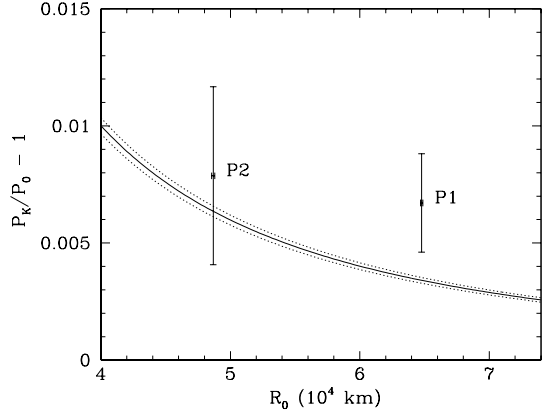}
\caption{
Deviation of azimuthal period $P_0$ from Keplerian orbital period
$P_K$ as a function of the guiding center radius $R_0$.
The solid line shows $P_K/P_0 - 1 = n_0/n_K - 1$ from Eq.
(\ref{n0}) for the best-fit mass ratio of Charon-Pluto, $m_c/m_p =
0.1165$, while the dotted lines show the same quantity for $m_c/m_p$
that are $1\,\sigma$ ($\pm 0.0055$) from the best-fit value.
The values of $P_K/P - 1$ and $a$ for the satellites P2 and P1, with
orbital periods $P$ and semimajor axes $a$ from unperturbed Keplerian
fits (Table \ref{table1}), are shown with their $1\,\sigma$ error
bars.
\label{fig:period}}
\end{figure}

We have analyzed the orbits of the recently discovered satellites of
Pluto, S/2005 P1 and S/2005 P2.
Because of the rather large mass ratio of Charon-Pluto ($m_c/m_p \sim
0.1$), the orbits of P2 and P1 are non-Keplerian even if P2 and P1
have negligible masses.
The analytic theory in Section 2 with P2 and P1 treated as test
particles shows that the motion in $R$ and $\phi$ can be described by
the superposition of the circular motion of the guiding center at mean
motion $n_0$, the epicyclic motion represented by eccentricity $e$ at
epicyclic frequency $\kappa_0$, and the forced oscillations at
frequencies $k|n_0-n_{pc}|$ ($k = 1,2,3,\ldots$) due to the
non-axisymmetric components of the gravitational potential rotating at
the mean motion $n_{pc}$ of Pluto-Charon;
the motion in $z$ at the vertical frequency $\nu_0$ decouples from
that in $R$ and $\phi$.
With $\nu_0 > n_0 > n_K > \kappa_0$, where $n_K$ is the Keplerian mean
motion about the center of mass of Pluto and Charon, we found that the
azimuthal period $P_0 = 2\pi/n_0$ is shorter than the Keplerian
orbital period and that the periapse and ascending node (relative to
the Pluto-Charon orbital plane) precess at nearly equal rates in
opposite directions (prograde for the periapse and retrograde for the
node).
We have also performed a series of direct numerical orbit integrations
with different assumed masses for P2 and P1, and the results presented
in Section 3 show the increasing effects of the proximity of the
orbits of P2 and P1 to the 3:2 mean-motion commensurability with
increasing masses.
As shown in Fig. \ref{fig:period}, the deviation from Kepler's third
law is already detected in the unperturbed Keplerian fit of BGYYS
(which was previously pointed out by BGYYS as discrepancies in the
total mass of Pluto-Charon inferred from the orbits of Charon, P2, and
P1).
Since the other non-Keplerian behaviors depend on the masses of P2 and
P1, a dynamical fit to the data that accounts for the interactions among
Charon, P2, and P1 should allow us to place constraints on the masses
of P2 and P1, although the existing data consisting of only 12
observations over a 1-year time span may not be sufficient.
(BGYYS included in their unperturbed Keplerian fits the discovery data
for P2 and P1 from \citealt{wea06}, which were taken two years after
the last observation of BGYYS, but the discovery data have larger
errors and may not be useful in constraining parameters other than the
orbital periods.)

\begin{figure}
\epsscale{0.54}
\plotone{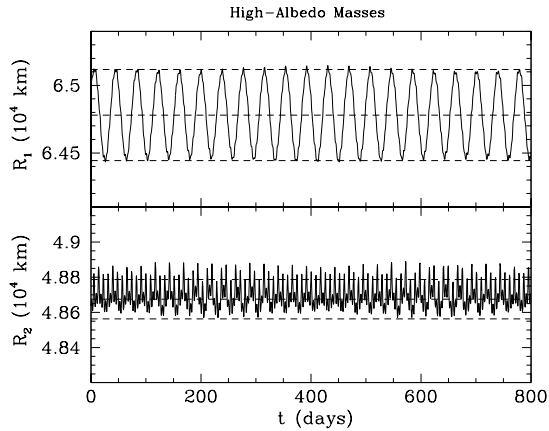}
\caption{
Variations in the orbital radii $R_2$ and $R_1$ of P2 and P1,
respectively, for $800\,$days in the numerical orbit integration with
the high-albedo masses for P2 and P1.
The dashed lines indicate the semimajor axes, $a$, and the maximum and
minimum radii, $a (1 \pm e)$, for the best-fit Keplerian orbits in
Table \ref{table1}.
The variation in $R_2$ is dominated by high-frequency forced
oscillations due to the non-axisymmetric components of the potential
rotating at the mean motion of Pluto-Charon, but a small epicyclic
motion with period $\approx 25.2\,$days is also visible.
The variation in $R_1$ is dominated by the epicyclic motion with
eccentricity $\approx 0.0052$ and period $\approx 38.6\,$days.
\label{fig:highR}}
\end{figure}

\begin{figure}
\epsscale{0.54}
\plotone{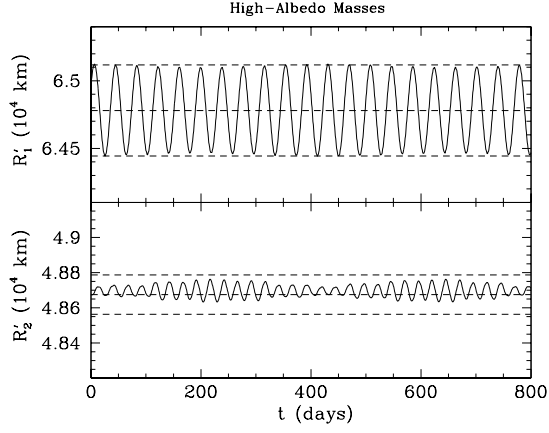}
\caption{
Same as Fig. \ref{fig:highR}, but for the transformed orbital radii
$R_2'$ and $R_1'$ (Eq. [\ref{Rprime}]).
Most of the forced oscillations in $R_2$ and $R_1$ are eliminated, and
small periodic variations in the maximum and minimum values of $R_2'$
and $R_1'$, or equivalently in the epicyclic eccentricities $e_2$ and
$e_1$, become visible.
The periods are about $400\,$days for $e_2$ and $450\,$days for $e_1$.
\label{fig:highRprime}}
\end{figure}

\begin{figure}
\epsscale{0.54}
\plotone{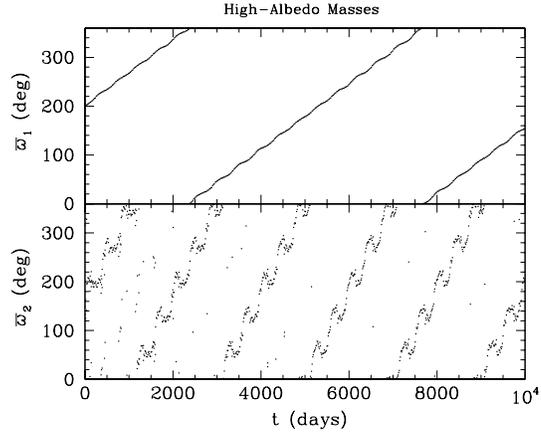}
\caption{
Evolution of the longitudes of periapse $\varpi_2$ and $\varpi_1$ of
P2 and P1, respectively, for $10^4\,$days in the high albedo case.
There are short-term periodic variations on the same timescales as
$e_2$ and $e_1$, but the long-term evolution is prograde precessions
with periods of $2000\,$days for P2 and $5300\,$days for P1.
\label{fig:highvarpi}}
\end{figure}

\begin{figure}
\epsscale{0.54}
\plotone{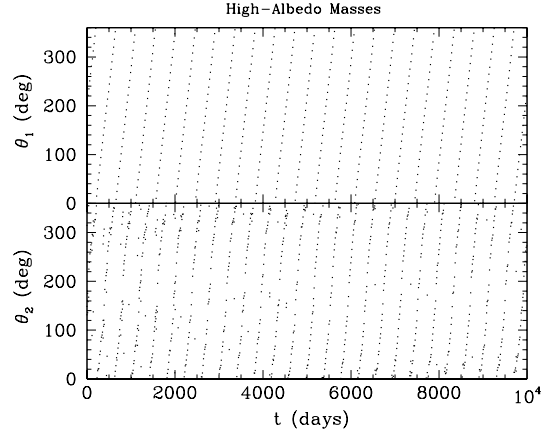}
\caption{
Same as Fig. \ref{fig:highvarpi}, but for the resonance variables
$\theta_2 = 2\phi_2 - 3\phi_1 + \varpi_2$ and  $\theta_1 = 2\phi_2 -
3\phi_1 + \varpi_1$ at the 3:2 commensurability between P2 and P1.
The period of the short-term variations in $e_2$ and $\varpi_2$ ($e_1$
and $\varpi_1$) is the circulation period of $\theta_2$ ($\theta_1$).
\label{fig:highresvar}}
\end{figure}

If the albedos of P2 and P1 are high and of order of that of Charon,
the masses of P2 and P1 are sufficiently low that their orbits are
well described by the analytic theory.
The largest correction due to the proximity to the 3:2
commensurability is a significant variation of the precession of
$\varpi_2$ on the period of circulation ($\approx 400\,$days) of the
resonance variable $\theta_2 = 2\phi_2 - 3\phi_1 + \varpi_2$
(Figs. \ref{fig:highvarpi} and \ref{fig:highresvar}).
However, the variation in the orbital radius $R_2$ of P2 due to the
forced oscillations are large (Fig. \ref{fig:highR}), and the
eccentricity with a large error ($0.0023 \pm 0.0021$) found by the
Keplerian fit of BGYYS probably results from an attempt to fit the 12
data points without taking into account the forced oscillations.
Thus, there is at present no evidence that P2 has any significant
epicyclic eccentricity, and it is likely that the periapse precession
of P2 would be difficult to measure.
On the other hand, the orbit of P1 has a significant epicyclic
eccentricity.
Over the 1-year time span of the existing data, the prograde
precession of $\varpi_1$ with a period of $5300\,$days
(Fig. \ref{fig:highvarpi}) would have resulted in a $25^\circ$ change
in $\varpi_1$, which may be difficult to detect with only 12 data
points.
But there would be more than $3.6\yr$ between the first existing data
point and any additional data of comparable quality taken after the
writing of this paper (Feb. 2006), and the $\ga 90^\circ$ precession
of $\varpi_1$ should be detectable.

\begin{figure}
\epsscale{0.54}
\plotone{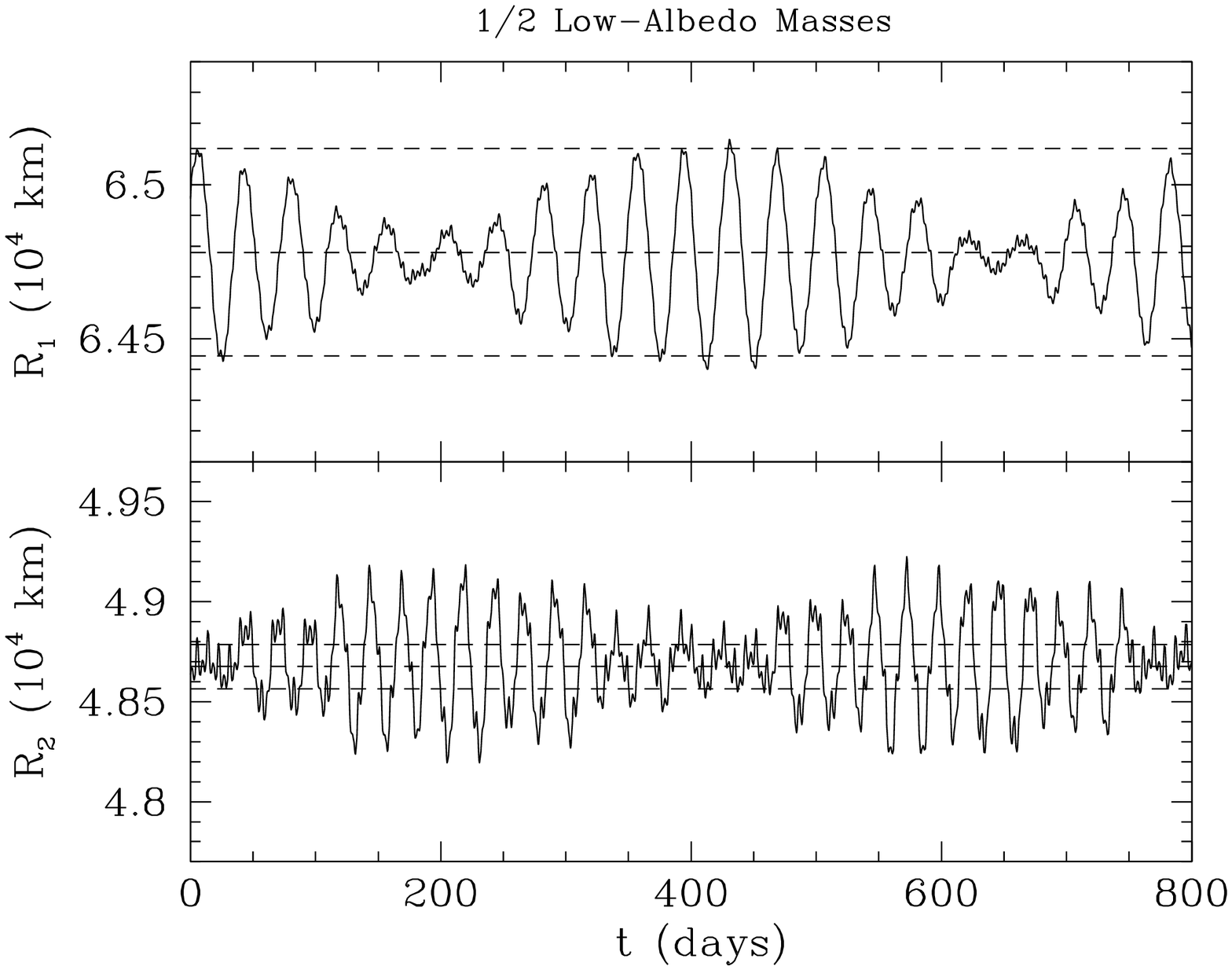}
\caption{
Same as Fig. \ref{fig:highR}, but for the numerical integration with
P2 and P1 having half the low-albedo masses.
The periodic variations in $e_2$ and $e_1$ have amplitudes that are
significantly larger than in the high-albedo case
(Fig. \ref{fig:highRprime}) and are visible without transforming $R_2$
and $R_1$ to $R_2'$ and $R_1'$.
\label{fig:halflowR}}
\end{figure}

\begin{figure}
\epsscale{0.54}
\plotone{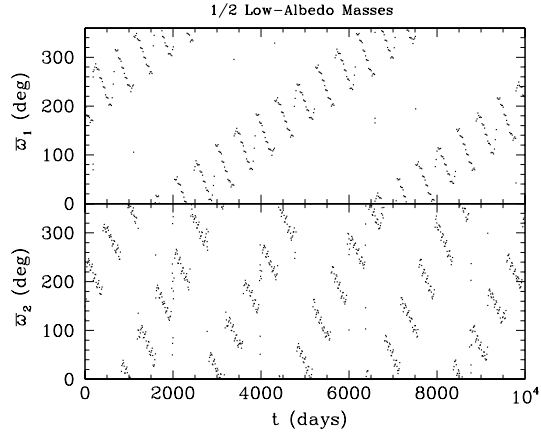}
\caption{
Same as Fig. \ref{fig:highvarpi}, but for the numerical integration
with P2 and P1 having half the low-albedo masses.
Compared to the high albedo case (Fig. \ref{fig:highvarpi}), the
amplitudes of the short-term periodic variations are significantly
larger, and the periods of the long-term prograde precession are
slightly shorter.
\label{fig:halflowvarpi}}
\end{figure}

\begin{figure}
\epsscale{0.54}
\plotone{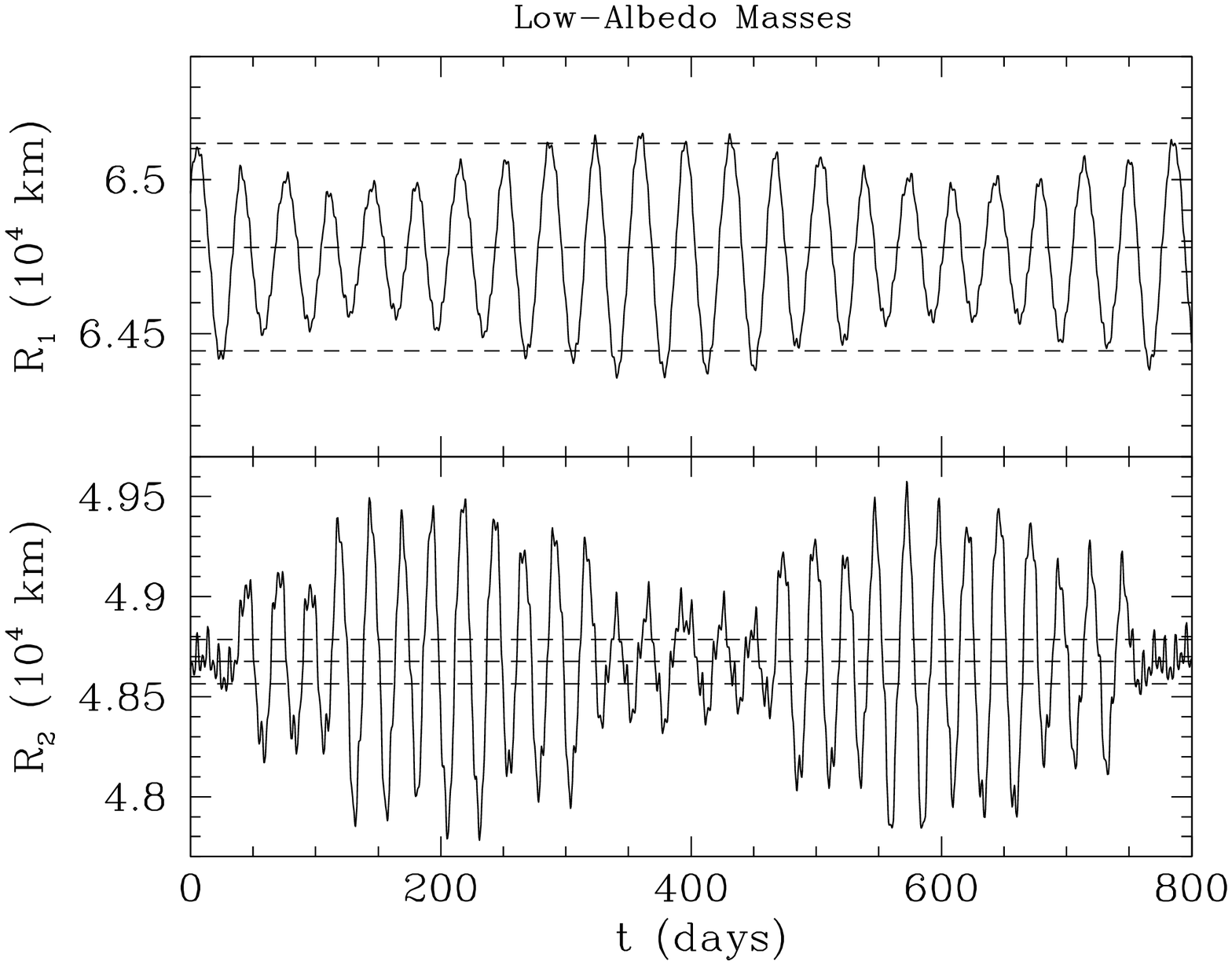}
\caption{
Same as Fig. \ref{fig:highR}, but for the numerical integration with
the low-albedo masses for P2 and P1.
The amplitude of the periodic variation in $e_2$ continues to increase
with mass, but that in $e_1$ is smaller than in the case with half the
low-albedo masses (Fig. \ref{fig:halflowR}).
\label{fig:lowR}}
\end{figure}

\begin{figure}
\epsscale{0.54}
\plotone{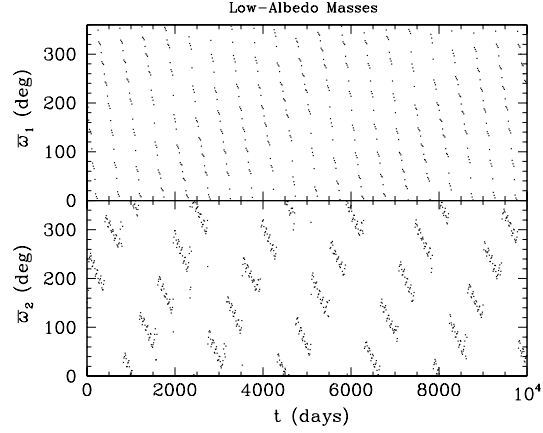}
\caption{
Same as Fig. \ref{fig:highvarpi}, but for the numerical integration
with the low-albedo masses for P2 and P1.
While $\varpi_2$ continues to show long-term prograde precession with
large short-term periodic variation, $\varpi_1$ shows a completely
retrograde precession with a period of $500\,$days.
\label{fig:lowvarpi}}
\end{figure}

\begin{figure}
\epsscale{0.54}
\plotone{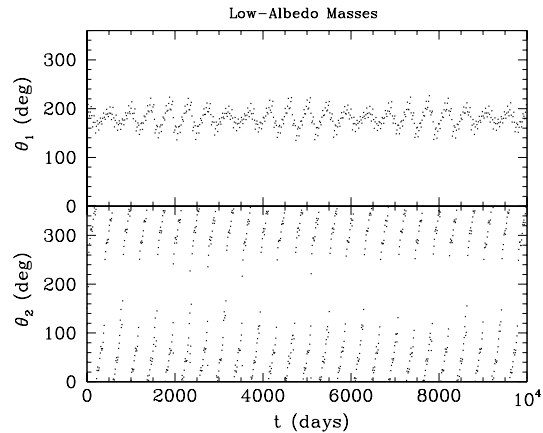}
\caption{
Same as Fig. \ref{fig:highresvar}, but for the numerical integration
with the low-albedo masses for P2 and P1.
While $\theta_2$ circulates as in the high albedo case shown in Fig.
\ref{fig:highresvar} (the nearly empty region about $180^\circ$ is due
to the fast prograde precession of $\varpi_2$ when $e_2$ is small),
$\theta_1$ librates about $180^\circ$.
\label{fig:lowresvar}}
\end{figure}

\begin{figure}
\epsscale{0.54}
\plotone{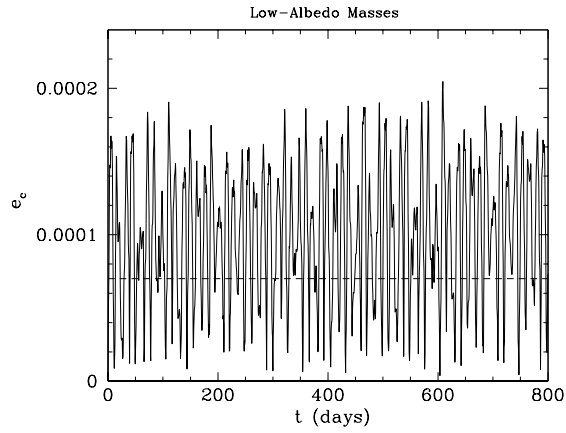}
\caption{
Variation in the orbital eccentricity $e_c$ of Charon for $800\,$days
in the numerical integration with the low-albedo masses for P2 and P1.
The dashed line indicates $e_c = 7.0 \times 10^{-5}$, which is
$1\,\sigma$ above the best-fit value ($e_c = 0.0$) in Table
\ref{table1}.
\label{fig:ec}}
\end{figure}

If the albedos of P2 and P1 are low and of order of that of comets,
the masses of P2 and P1 are sufficiently large that there are
significant short-term variations in their epicyclic eccentricities
and/or periapse longitudes due to the proximity to the 3:2
commensurability.
In the case with half the low-albedo masses (Figs. \ref{fig:halflowR}
and \ref{fig:halflowvarpi}), there are significant variations in $e_2$
and $\varpi_2$ on the circulation period of $\approx 400\,$days of
$\theta_2$, and in $e_1$ and $\varpi_1$ on the circulation period of
$\approx 450\,$days of $\theta_1 = 2\phi_2 - 3\phi_1 + \varpi_1$.
In the case with low-albedo masses (Figs. \ref{fig:lowR},
\ref{fig:lowvarpi}, and \ref{fig:lowresvar}), $\theta_1$ is in
resonance and $\varpi_1$ shows a retrograde precession with a period
of only $500\,$days.
Since the existing data can be reasonably fitted by unperturbed
Keplerian orbits, one might speculate that the existing data are
already inconsistent with masses of P2 and P1 near the upper end of
the expected range.
But we cannot rule out the possibility that orbits like those in
Figs. \ref{fig:halflowR}--\ref{fig:lowvarpi} can be fitted by
Keplerian orbits, if the orbits are sparsely sampled as in the
existing data set.
Another indication that masses of P2 and P1 near the upper end of the
expected range may already be ruled out comes from the orbital
eccentricity, $e_c$, of Charon.
As pointed out by \citet{ste94}, the perturbations from additional
satellites in the Pluto-Charon system induce eccentricity in Charon's
orbit, and the observed value or upper limit of $e_c$ can be used to
constrain the masses of the additional satellites.
In Fig. \ref{fig:ec} we show the variation of $e_c$ for $800\,$days
in the run with low-albedo masses.
The eccentricity $e_c$ varies up to $2 \times 10^{-4}$, which is
significantly larger than the best-fit $e_c = 0.0 \pm 7.0 \times
10^{-5}$ in Table \ref{table1}.
But it should be noted that the random error of $7.0 \times 10^{-5}$
corresponds to shifts of the order of 0.1 mas in the positions of
Charon relative to Pluto, and it is unclear that systematic error in,
e.g., the correction from the center of light to the center of body is
below 0.1 mas.
In any case, a data set that samples more densely any possible
variations in the orbits of P2 and P1 on the $400$-$500$-day
timescales should provide strong constraints on the masses of P2 and
P1.

There are no apparent effects due to the respective proximity of P2
and P1 to the high-order 4:1 and 6:1 mean-motion commensurabilities
with Charon.
However, \citet{war06} have proposed that P2 and P1 may have been
trapped in the corotation resonances at these commensurabilities
(i.e., those associated with the resonance variable $\phi_c - 4\phi_2
+ 3\varpi_c$ for P2 and $\phi_c - 6\phi_1 + 5\varpi_c$ for P1) during
the tidal expansion of Charon's orbit, if Charon formed with large
orbital eccentricity from a giant impact on Pluto.
P2 and P1 escaped from the corotation resonances when Charon's orbital
eccentricity was tidally damped to very small value.

Our analysis shows that continued observations of the Pluto system
with HST (and possibly ground-based adaptive optics) in the near
future will allow us to detect the non-Keplerian behaviors of the
orbits of P2 and P1 (in addition to the already detected deviation
from Kepler's third law) and to thereby constrain their masses.
Much more precise determination of the orbits and masses of P2 and P1
will be possible as the New Horizons spacecraft approaches the Pluto
system in 2015.

\acknowledgments
We thank Robin Canup and Alan Stern for sending us preprints on the
new satellites of Pluto.
This research was supported in part by NASA grant NNG05GK58G.

\clearpage

\begin{deluxetable}{lccc}
\tablecolumns{4}
\tablewidth{0pt}
\tablecaption{Orbital Parameters at Epoch JD 2452600.5 from Keplerian
              fits by \cite{bui06}
\label{table1}}
\tablehead{
\colhead{Parameter} & \colhead{Charon} & \colhead{S/2005 P2} &
\colhead{S/2005 P1}
}
\startdata
Period $P$ (days)                   & 6.3872304(11) & 24.8562(13) & 38.2065(14)\\
Semimajor Axis $a$ (km)             & 19571.4(4.0)  & 48675(121)  & 64780(88)\\
Eccentricity $e$                    & 0.000000(70)  & 0.0023(21)  & 0.0052(11)\\
Inclination $i$ (deg)               & 96.145(14)    & 96.18(22)   & 96.36(12)\\
Long. Ascending Node $\Omega$ (deg) & 223.046(14)   & 223.14(23)  & 223.173(86)\\
Long. Periapse $\varpi$ (deg)       & \nodata       & 216(13)     & 200.1(3.7)\\
Mean Long. at Epoch $\lambda$ (deg) & 257.946(13)   & 123.14(20)  & 322.71(23)\\
\enddata
\tablecomments{The parameters are for the orbit of Charon relative to
Pluto and the orbits of P1 and P2 relative to the center of mass of
Pluto-Charon.
Numbers in parentheses are $1\,\sigma$ errors in the least significant
digits.}
\end{deluxetable}

\clearpage

\begin{deluxetable}{lcc}
\tablecolumns{3}
\tablewidth{0pt}
\tablecaption{Parameters of Analytic Theory
\label{table2}}
\tablehead{
\colhead{Parameter} & \colhead{S/2005 P2} & \colhead{S/2005 P1}
}
\startdata
$R_0$ (km)              & 48675   & 64780\\
$P_K = 2\pi/n_K$ (days) & 25.0518 & 38.4628\\
$n_0/n_K$               & 1.00635 & 1.00341\\
$\kappa_0/n_K$          & 0.99198 & 0.99612\\
$\nu_0/n_K$             & 1.02053 & 1.01063\\
$C_1$                   & 0.001275 & 0.000149\\
$C_2$                   & 0.001373 & 0.000228\\
$C_3$                   & 0.000204 & 0.000026\\
$C_4$                   & 0.000044 & 0.000004\\
$D_1$                   & 0.003220 & 0.000458\\
$D_2$                   & 0.006813 & 0.001764\\
$D_3$                   & 0.001496 & 0.000314\\
$D_4$                   & 0.000437 & 0.000072\\
\enddata
\end{deluxetable}

\clearpage

\clearpage

\clearpage

\clearpage

\clearpage

\clearpage

\end{document}